# Exact Decomposition of Multifrequency Discrete Real and Complex Signals

BaoGuo Liu[*]

School of Mechanical and Electrical Engineering, Henan University of Technology;
Henan Key Laboratory of Super Hard Abrasives Grinding Equipment, Zhengzhou 45001, Henan, China

**Abstract**

'The spectral leakage from windowing and the picket fence effect from discretization' have been among the standard contents in textbooks for many decades. The spectral leakage and picket fence effect would cause the distortions in amplitude, frequency, and phase of signals, which have always been of concern, and attempts have been made to solve them. This paper proposes two novel decomposition theorems that can totally eliminate the spectral leakage and picket fence effect, and could broaden the knowledge of signal processing. First, two generalized eigenvalue equations are constructed for multifrequency discrete real signals and complex signals. The two decomposition theorems are then proved. On these bases, exact decomposition methods for real and complex signals are proposed. For a noise-free multifrequency real signal with $m$ sinusoidal components, the frequency, amplitude, and phase of each component can be exactly calculated by using just $4m-1$ discrete values and its second-order derivatives. For a multifrequency complex signal, only $2m-1$ discrete values and its first-order derivatives are needed. The numerical experiments show that the proposed methods have very high resolution, and the sampling rate does not necessarily obey the Nyquist sampling theorem. With noisy signals, the proposed methods have extraordinary accuracy.

**Keywords:** Exact decomposition; Multifrequency sinusoidal signal; Discrete real signal; Discrete complex signal; Decomposition theorem; Generalized eigenvalue equation

## 1. Introduction

Since Cooley and Tukey proposed the fast Fourier transform (FFT) of Fourier analysis in 1965 [1, 2], the technique has been indispensable in electronics, communication, signal analysis, digital image and audio processing, and many other fields [2, 3]. However, while performing the FFT algorithm, distortions in amplitude, frequency, and phase — caused by spectrum leakage (SL) due to signal truncation and the picket fence effect (PEF) due to frequency discretization — are inevitable, which have always been of concern, and attempts have been made to solve them.

In 1970, Rife and Vincent studied the correction measurement of frequencies and levels of tones using discrete Fourier transform (DFT) [4]. A number of scholars have also studied this issue [4-18]. Interpolation techniques are the most studied and applied methods in many engineering fields [4–10]. Interpolation techniques refer to interpolated discrete Fourier transform (IpDFT) or interpolated fast Fourier transform (IpFFT). The IpDFT and IpFFT reduce SL and PEF through windowing and interpolation, respectively. Generally, windows with the maximum side lobe decay efficiently reduce SL; thus, the Hanning window is one of the commonly used [5, 7, 8]. According to the distribution of spectrum lines in the windows, weighting the discrete signals before DFT and FFT are performed, and thus the PEF is decreased to a minimum, and accurate amplitudes, frequencies, and phases of the signal components can be computed [4–10]. IpDFT and IpFFT can be quickly and easily implemented; however, their results are not accurate when sinusoids are not well separated in frequency [5–8]. Weighted phase average (WPA) is another popular choice of a correction technique for FFT [11–15]. Relying on weighted linear regression on the phases, WPA averages the weighted phase estimators that are obtained by FFT within different nonoverlapping segments of the signals. WPA has the advantage of accurately calculated frequencies and phases, but amplitudes are dependent on the windows. There are many other correction techniques, such as the phase difference correction method [16] and sliding-window DFT method [17, 18, 27]. All of the correction techniques have advantages and disadvantages.

Distinct from techniques based on DFT and FFT, matrix-based singular value decomposition (SVD) and singular spectral decomposition (SSD) are techniques that aim to represent signals as a linear superposition of elementary variable modes with unnecessary harmonic components [19–22]. The aforementioned techniques do not provide estimators for the spectrum, but they are powerful denoising filters capable of separating autocoherent features, such as anharmonic oscillations and quasiperiodic phenomena, from random features. They are non-parametric techniques.

Other techniques commonly used to extract signal features include the wavelet transform (WT) [23, 24], Hilbert-Huang transform (HHT) [25], estimation of signal parameters via rotational invariance technique (ESPRIT) [26, 27], and multiple signal classification (MUSIC) [28, 29]. However, it is impossible to exactly compute frequencies, amplitudes, and phases of multifrequency discrete signals using the aforementioned techniques.

This paper constructs the generalized eigenvalue equations and proves the decomposition theorems for multifrequency discrete real signals and complex signals. Based on the two decomposition theorems, decomposition methods for real and complex signals are proposed. For a noise-free real signal with *m* sinusoidal components, the frequency, amplitude, and phase of each component can be exactly calculated by using just 4*m*–1 discrete values and its second-order derivatives. For a complex signal, only 2*m*–1 discrete values and its first-order derivatives are needed. The decomposition results are exact in theory. The numerical experiments show that the proposed methods have very high resolution, and the sampling rate does not necessarily obey the Nyquist sampling theorem. With noisy signals, the proposed methods have extraordinary accuracy.

## 2. Decomposition theorem of multifrequency discrete real signal

Multifrequency sinusoidal signal is one of the most common signals in engineering. The Multifrequency real signal can be presented as follows

$$x(t) = \sum_{i=1}^{m} a_i \sin(2\pi f_i t + \varphi_{i0}) \tag{1}$$

where $a_i$, $f_i$, and $\varphi_{i0}$ are the amplitude, frequency and phase of the *ith* component of signal $x(t)$; *m* is the number of component signals.

Discretize $x(t)$, and construct the following generalized eigenvalue equation of the multifrequency discrete real sinusoidal signal

$$-\boldsymbol{D}_x \boldsymbol{v} = \lambda \boldsymbol{X} \boldsymbol{v} \tag{2}$$

where $\lambda$ is the generalized eigenvalue; $\boldsymbol{v}$ is the generalized eigenvector; $\boldsymbol{X}$ is the square Hankel matrix of $x(t)$; and $\boldsymbol{D}_x$ is the square Hankel matrix of the second-order derivative $\ddot{x}(t)$. And

$$\boldsymbol{X} = \begin{bmatrix} x_1 & x_2 & \cdots & x_n \\ x_2 & x_3 & \cdots & x_{n+1} \\ \vdots & \vdots & \ddots & \vdots \\ x_n & x_{n+1} & \cdots & x_{2n-1} \end{bmatrix}; \quad \boldsymbol{D}_x = \begin{bmatrix} \ddot{x}_1 & \ddot{x}_2 & \cdots & \ddot{x}_n \\ \ddot{x}_2 & \ddot{x}_3 & \cdots & \ddot{x}_{n+1} \\ \vdots & \vdots & \ddots & \vdots \\ \ddot{x}_n & \ddot{x}_{n+1} & \cdots & \ddot{x}_{2n-1} \end{bmatrix} \tag{3}$$

where $x_k, \ddot{x}_k$ ($k = 1, \cdots, 2n-1$) are the discrete series of $x(t)$ and $\ddot{x}(t)$, respectively.

**Theorem 1:** When $n \geq 2m$, the generalized eigenvalue equation of the multifrequency real sinusoidal signal has $2m$ nonzero generalized eigenvalues, which are given by the following equations

$$\lambda_{2i-1} = \lambda_{2i} = 4\pi^2 f_i^2, \quad (i = 1, \cdots, m) \tag{4}$$

## 2.1 Proof

The square Hankel matrix $\boldsymbol{X}$ of signal $x(t)$ is rewritten as follows

$$\boldsymbol{X} = \sum_{i=1}^{m} a_i \boldsymbol{X}_i \tag{5}$$

where $a_i \boldsymbol{X}_i$ is the square Hankel matrix of the $ith$ sinusoidal component. The $kth$ row and $lth$ column element of matrix $\boldsymbol{X}_i$ is given by the following equation

$$x_{kl}^i = \sin(2\pi f_i(k+l-2)/f_s + \varphi_{i0}), \quad (k,l = 1, \cdots, n) \tag{6}$$

were $f_s$ is sampling rate. The square Hankel matrix $\boldsymbol{D}_x$ is given by the following equation

$$\boldsymbol{D}_x = \sum_{i=1}^{m} \boldsymbol{D}_{xi} = \sum_{i=1}^{m} -4\pi^2 f_i^2 a_i \boldsymbol{X}_i \tag{7}$$

The ranks of matrices $\boldsymbol{X}$ and $\boldsymbol{D}_x$ are less than or equal to $2m$, because the ranks of the square Hankel matrices $\boldsymbol{X}_i$ and $\boldsymbol{D}_{x_i}$ are two [30].

Suppose $\lambda = 4\pi^2 f_i^2$, and considering the relation $\boldsymbol{D}_{xi} = -4\pi^2 f_i^2 a_i \boldsymbol{X}_i$, eq. (2) is modified to obtain the following equation

$$\left\{ \sum_{k=1, k \neq i}^{m} \boldsymbol{D}_{xk} + 4\pi^2 f_i^2 \sum_{k=1, k \neq i}^{m} a_k \boldsymbol{X}_k \right\} \boldsymbol{v} = \boldsymbol{0} \tag{8}$$

Eq. (8) has at least two nonzero solutions, which are $\boldsymbol{v}_{2i}$ and $\boldsymbol{v}_{2i-1}$, because it is a homogeneous equation about vector $\boldsymbol{v}$, and the rank of the coefficient matrix is less than or equal to $2m - 2$.

In eq. (8), replacing $\boldsymbol{v}$ with $\boldsymbol{v}_{2i}$, and left multiplying by $\boldsymbol{v}_{2i}^T$, yields

$$4\pi^2 f_i^2 = -\boldsymbol{v}_{2i}^T \left( \sum_{k=1, k \neq i}^{m} \boldsymbol{D}_{xk} \right) \boldsymbol{v}_{2i} \bigg/ \boldsymbol{v}_{2i}^T \left( \sum_{k=1, k \neq i}^{m} a_k \boldsymbol{X}_k \right) \boldsymbol{v}_{2i} \tag{9}$$

where $\boldsymbol{v}_{2i}^T$ is the transpose vector of $\boldsymbol{v}_{2i}$. Considering the following identical equation

$$4\pi^2 f_i^2 = -\frac{\boldsymbol{v}_{2i}^T \boldsymbol{D}_{xi} \boldsymbol{v}_{2i}}{\boldsymbol{v}_{2i}^T a_i \boldsymbol{X}_i \boldsymbol{v}_{2i}} \tag{10}$$

The following equation is derived from eq. (9) and (10).

$$4\pi^2 f_i^2 = \frac{-\boldsymbol{v}_{2i}^T \boldsymbol{D}_x \boldsymbol{v}_{2i}}{\boldsymbol{v}_{2i}^T \boldsymbol{X} \boldsymbol{v}_{2i}} \tag{11}$$

Eq. (11) reveals that $4\pi^2 f_i^2$ is a Rayleigh quotient of matrices $-\boldsymbol{D}_x$ and $\boldsymbol{X}$, that is, $\lambda = 4\pi^2 f_i^2$ is a generalized eigenvalue of eq. (2), and $\boldsymbol{v}_{2i}$ is the corresponding generalized eigenvector [29]. Replacing $\boldsymbol{v}_{2i}$ with $\boldsymbol{v}_{2i-1}$, eq. (11) is also established. Therefore, $\lambda = 4\pi^2 f_i^2$ is a double eigenvalue of eq. (2).

Decomposition theorem 1 of the discrete real signal is thus proven.

### 2.2 Component amplitudes and phases of multifrequency discrete real signal

After the $i$th double eigenvalue and its two corresponding eigenvectors are computed, considering eq. (6), $\boldsymbol{v}_{2i}^T a_i \boldsymbol{X}_i \boldsymbol{v}_{2i}$ and $\boldsymbol{v}_{2i-1}^T a_i \boldsymbol{X}_i \boldsymbol{v}_{2i-1}$ can be expanded as follows

$$\begin{cases} \boldsymbol{v}_{2i}^T a_i \boldsymbol{X}_i \boldsymbol{v}_{2i} = \boldsymbol{v}_{2i}^T \boldsymbol{H}_{i1} \boldsymbol{v}_{2i} a_i \cos\varphi_{i0} + \boldsymbol{v}_{2i}^T \boldsymbol{H}_{i2} \boldsymbol{v}_{2i} a_i \sin\varphi_{i0} \\ \boldsymbol{v}_{2i-1}^T a_i \boldsymbol{X}_i \boldsymbol{v}_{2i-1} = \boldsymbol{v}_{2i-1}^T \boldsymbol{H}_{i1} \boldsymbol{v}_{2i-1} a_i \cos\varphi_{i0} + \boldsymbol{v}_{2i-1}^T \boldsymbol{H}_{i2} \boldsymbol{v}_{2i-1} a_i \sin\varphi_{i0} \end{cases} \tag{12}$$

where $\boldsymbol{H}_{i1}$ and $\boldsymbol{H}_{i2}$ are square matrices with dimensions $n$. Their $k$th row and $l$th column elements are given by the following equation

$$\begin{cases} \{h\}_{kl}^{i1} = \sin(2\pi f_i(k+l-2)/f_s) \\ \{h\}_{kl}^{i2} = \cos(2\pi f_j(k+l-2)/f_s) \end{cases}, (k, l = 1, \cdots n) \tag{13}$$

For the $i$th generalized eigenvector, $\boldsymbol{v}_{2i}^T a_i \boldsymbol{X}_i \boldsymbol{v}_{2i}$ can be replaced with $\boldsymbol{v}_{2i}^T \boldsymbol{X} \boldsymbol{v}_{2i}$, and $\boldsymbol{v}_{2i-1}^T a_i \boldsymbol{X}_i \boldsymbol{v}_{2i-1}$ can be replaced by $\boldsymbol{v}_{2i-1}^T \boldsymbol{X} \boldsymbol{v}_{2i-1}$, because the generalized eigenvectors are weighted orthogonally, that is, $\boldsymbol{v}_{2i}^T \boldsymbol{X} \boldsymbol{v}_k = \boldsymbol{v}_{2i-1}^T \boldsymbol{X} \boldsymbol{v}_k = 0$ ($k \neq 2i, k \neq 2i-1$). Thus eq. (12) is modified to create

$$\begin{bmatrix} \boldsymbol{v}_{2i}^T \boldsymbol{H}_{i1} \boldsymbol{v}_{2i} & \boldsymbol{v}_{2i}^T \boldsymbol{H}_{i2} \boldsymbol{v}_{2i} \\ \boldsymbol{v}_{2i-1}^T \boldsymbol{H}_{i1} \boldsymbol{v}_{2i-1} & \boldsymbol{v}_{2i-1}^T \boldsymbol{H}_{i2} \boldsymbol{v}_{2i-1} \end{bmatrix} \begin{Bmatrix} a_i \cos\varphi_{i0} \\ a_i \sin\varphi_{i0} \end{Bmatrix} = \begin{Bmatrix} \boldsymbol{v}_{2i}^T \boldsymbol{X} \boldsymbol{v}_{2i} \\ \boldsymbol{v}_{2i-1}^T \boldsymbol{X} \boldsymbol{v}_{2i-1} \end{Bmatrix} \tag{14}$$

The amplitude $a_i$ and phase $\varphi_{i0}$ can be calculated after the unknowns $a_i \cos\varphi_{i0}$ and $a_i \sin\varphi_{i0}$ are computed from eq. (14). The frequencies, amplitudes, and phases of all sinusoidal component signals can be computed in this way.

## 3. Decomposition theorem of multifrequency discrete complex signal

The multifrequency complex signal can be expressed as follows

$$y(t) = \sum_{i=1}^{m} a_i e^{j(2\pi f_i t + \varphi_{i0})} \tag{15}$$

where $a_i$, $f_i$, and $\varphi_{i0}$ are the amplitude, frequency and phase of the $i$th component of complex signal $y(t)$; $m$ is the number of component complex signals.

Discretize $y(t)$, and construct the generalized eigenvalue equation of the multifrequency discrete complex signal as follows

$$-j\boldsymbol{D}_y \boldsymbol{v} = \lambda \boldsymbol{Y} \boldsymbol{v} \tag{16}$$

where $\lambda$ is the generalized eigenvalue; $\boldsymbol{v}$ is the generalized eigenvector; $\boldsymbol{Y}$ is the square Hankel matrix of $y(t)$; and $\boldsymbol{D}_y$ is the square Hankel matrix of the first-order derivative $\dot{y}(t)$. And

$$Y = \begin{bmatrix} y_1 & y_2 & \cdots & y_n \\ y_2 & y_3 & \cdots & y_{n+1} \\ \vdots & \vdots & \ddots & \vdots \\ y_n & y_{n+1} & \cdots & y_{2n-1} \end{bmatrix}; \quad D_y = \begin{bmatrix} \dot{y}_1 & \dot{y}_2 & \cdots & \dot{y}_n \\ \dot{y}_2 & \dot{y}_3 & \cdots & \dot{y}_{n+1} \\ \vdots & \vdots & \ddots & \vdots \\ \dot{y}_n & \dot{y}_{n+1} & \cdots & \dot{y}_{2n-1} \end{bmatrix} \quad (17)$$

where $y_k, \dot{y}_k$ ($k = 1, \cdots, 2n-1$) are the discrete series of $y(t)$ and $\dot{y}(t)$, respectively.

**Theorem 2:** When $n \geq m$, the generalized eigenvalue equation of the multifrequency complex signal has $m$ nonzero generalized eigenvalues, which are given by the following equation

$$\lambda_i = 2\pi f_i, \quad (i = 1, \cdots, m) \quad (18)$$

### 3.1 Proof

The square Hankel matrix $Y$ of complex signal $y(t)$ can be rewritten as follows

$$Y = \sum_{i=1}^{m} a_i Y_i \quad (19)$$

where $a_i Y_i$ is the square Hankel matrix of the $ith$ component complex signal. The $kth$ row and $lth$ column element of matrix $Y_i$ is given by the following equation

$$y_{kl}^i = e^{j(2\pi f_i(k+l-2)/f_s + \varphi_{i0})}, \quad (k, l = 1, \cdots, n) \quad (20)$$

were $f_s$ is sampling rate. The square Hankel matrix $D_y$ is given by the following equation

$$D_y = \sum_{i=1}^{m} D_{yi} = \sum_{i=1}^{m} j 2\pi f_i a_i Y_i \quad (21)$$

The ranks of matrices $Y$ and $D_y$ are less than or equal to $m$, because the ranks of the square Hankel matrices $Y_i$ and $D_{y_i}$ are both one [30].

Suppose $\lambda = 2\pi f_i$, and considering the relation $D_{yi} = j 2\pi f_i a_i Y_i$, eq. (16) is modified to obtain the following equation

$$\left\{ j \sum_{k=1, k \neq i}^{m} D_{yk} + 2\pi f_i \sum_{k=1, k \neq i}^{m} a_k Y_k \right\} v = 0 \quad (22)$$

Eq. (22) has at least one nonzero solution, which is $v_i$, because it is a homogeneous equation about vector $v$, and the rank of the coefficient matrix is less than or equal to $m - 1$.

In eq. (22), replacing $v$ with $v_i$, and left multiplying by $v_i^T$, yields

$$2\pi f_i = -v_i^T \left( j \sum_{k=1, k \neq i}^{m} D_{yk} \right) v_i \bigg/ v_i^T \left( \sum_{k=1, k \neq i}^{m} a_k Y_k \right) v_i \quad (23)$$

where $v_i^T$ is the transpose vector of $v_i$. Considering the following identical equation

$$2\pi f_i = -\frac{j v_i^T D_{yi} v_i}{v_i^T a_i Y_i v_i} \quad (24)$$

The following equation is derived from eq. (23) and (24).

$$2\pi f_i = \frac{-v_i^T j D_y v_i}{v_i^T Y v_i} \quad (25)$$

Eq. (25) reveals that $2\pi f_i$ is a Rayleigh quotient of matrices $-j\boldsymbol{D}_y$ and $\boldsymbol{Y}$, that is, $\lambda = 2\pi f_i$ is a generalized eigenvalue of eq. (16), and $\boldsymbol{v}_i$ is the corresponding generalized eigenvector [29]. Therefore, $\lambda = 2\pi f_i$ is a eigenvalue of eq. (16).

Decomposition theorem 2 of the discrete complex signal is thus proven.

### 3.2 Component amplitudes and phases of multifrequency discrete complex signal

After the $ith$ eigenvalue and its corresponding eigenvector are computed, considering eq. (20), $\boldsymbol{v}_i^T a_i \boldsymbol{Y}_i \boldsymbol{v}_i$ can be expressed as follows

$$\boldsymbol{v}_i^T a_i \boldsymbol{Y}_i \boldsymbol{v}_i = a_i e^{j\varphi_{i0}} \boldsymbol{v}_i^T \boldsymbol{G}_i \boldsymbol{v}_2 \tag{26}$$

where $\boldsymbol{G}_i$ is a square matrix with dimensions $n$. Its $kth$ row and $lth$ column element is given by the following equation

$$\{g\}_{kl}^i = e^{j(2\pi f_i (k+l-2)/f_s)}, \quad (k, l = 1, \cdots n) \tag{27}$$

For the $ith$ generalized eigenvector, $\boldsymbol{v}_i^T a_i \boldsymbol{Y}_i \boldsymbol{v}_i$ can be replaced with $\boldsymbol{v}_i^T \boldsymbol{Y} \boldsymbol{v}_i$, because the generalized eigenvectors are weighted orthogonally, that is, $\boldsymbol{v}_i^T \boldsymbol{Y} \boldsymbol{v}_k = 0 \ (k \neq i)$. Thus eq. (26) is modified to create

$$a_i e^{j\varphi_{i0}} = \frac{\boldsymbol{v}_i^T \boldsymbol{Y} \boldsymbol{v}_i}{\boldsymbol{v}_i^T \boldsymbol{G}_i \boldsymbol{v}_i}, \quad (i = 1, \cdots m) \tag{28}$$

Eq. (28) is a complex equation, so the amplitude $a_i$ and phase $\varphi_{i0} \ (i = 1, \cdots m)$ can be calculated.

## 4. Numerical experiments

### 4.1 Decomposition of noise-free multifrequency real signal and complex signal

Based on eq. (1) and eq. (15) as the noise-free multifrequency real signal and complex signal, respectively, the known frequencies, amplitudes and phases of all component signals for $m$=10 are listed in Table 1. Since the highest frequency of the component signal is 290 $Hz$, we can set the sampling rate $f_s$=299 $Hz$, which is slightly higher than the highest frequency.

Table 1. Parameters of the known signal

| $f_j(Hz)$ | 35 | 60 | 97 | 97.5 | 120 | 135 | 160 | 230 | 270 | 290 |
|---|---|---|---|---|---|---|---|---|---|---|
| $a_j$ | 1.5 | 3.5 | 2 | 0.1 | 1.2 | 0.8 | 2.5 | 0.8 | 1 | 0.3 |
| $\varphi_{j0}(°)$ | 30 | 50 | 170 | 230 | 90 | 145 | 0 | 330 | 280 | 0 |

The minimum frequency difference between two neighboring components is only 0.5 $Hz$, the relative difference is about 0.5%, and the corresponding amplitude difference is up to 20 times as large.

For the real signal where $n$=4$m$−1=39, solve the generalized eigenvalue eq. (2), while ten double eigenvalues and twenty corresponding eigenvectors were computed. Ten component frequencies were calculated by eq. (4), and the component amplitudes and phases were computed using eq. (14).

Fig. 1 shows the known multifrequency real signal and its ten decomposed component signals. The 39 discrete points on the original multifrequency signal are identified by symbol 'o'.

Although the minimum frequency difference between two neighboring components is only 0.5 $Hz$, the relative difference is approximately 0.5%, and the corresponding amplitude difference is up to 20 times as large. The two neighboring component signals can be decomposed accurately. The sampling rate is less than the Nyquist sampling rate, which is 580 $Hz$ – twice the highest component frequency.

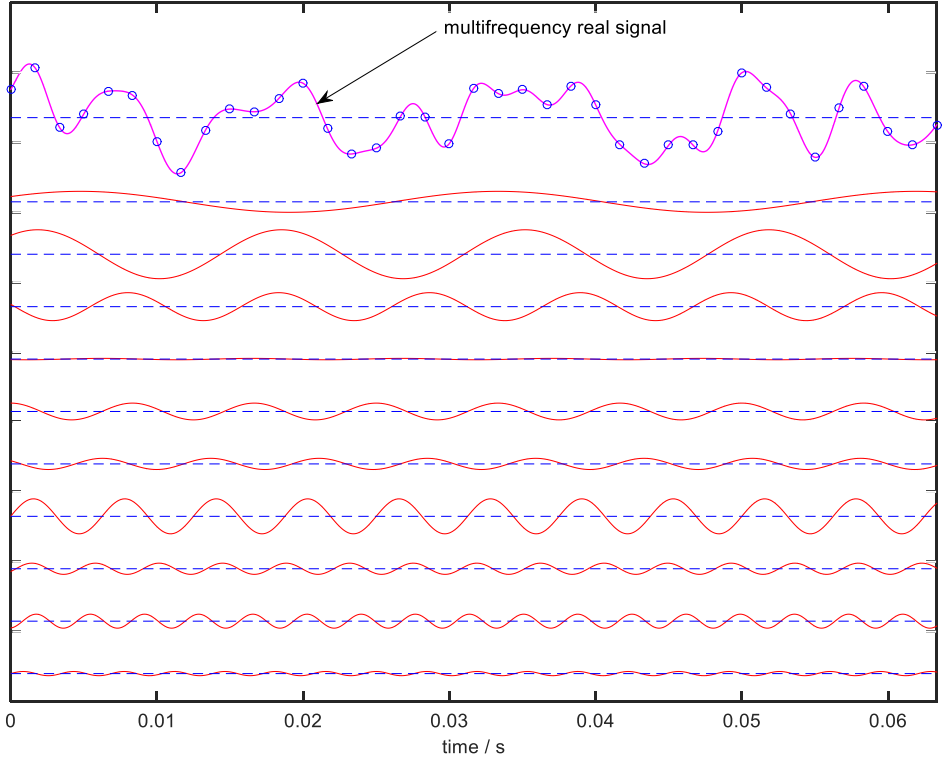

**Fig. 1.** Multifrequency real signal and its ten decomposed component signals. The 39 discrete points on the original multifrequency signal are identified by symbol 'o'. The maximum absolute errors in frequency, amplitude and phase are 1.99e-09, 7.79e-10 and -4.39e-08, respectively.

For the complex signal where $n=2m-1=19$, solve the generalized eigenvalue eq. (16), while ten eigenvalues and ten corresponding eigenvectors were computed. Ten component frequencies were calculated by eq. (18), and the component amplitudes and phases were computed using eq. (28).

The decomposed results are similar to those of the real signal. With the multifrequency complex signal, however, the decomposed results have both real component signals and imaginary component signals. The discrete points on the original signal are much less than the real signal.

The absolute computational errors in the component frequencies, amplitudes and phases are listed in Table 2. With the real signal, the maximum absolute errors in frequency, amplitude and phase are 1.99e-09, 7.79e-10 and -4.39e-08, respectively. With the complex signal, these errors are 6.41e-07, -2.51e-07 and -6.72e-06, respectively. It should be noted that all maximum errors resulted from the two neighboring components with minimum frequency difference and maximum amplitude difference, and the other errors were much less. Therefore, these errors were considered to be caused by the computer rather than the method.

Therefore, the proposed two decomposition theorems are exactly in theory. The corresponding decomposition methods are highly accurate and possess high resolution, and it is unnecessary for the sampling rate to obey the sampling theorem. In fact, it is impossible to extract accurate frequencies, amplitudes and phases from multifrequency signals by using other methods with so little discrete data.

Table 2. The absolute computation errors
in the component frequencies, amplitudes and phases

| *Errors of real signal* | | | *Errors of complex signal* | | |
|---|---|---|---|---|---|
| $\Delta f_j(Hz)$ | $\Delta a_j$ | $\Delta \varphi_{j0}(°)$ | $\Delta f_j(Hz)$ | $\Delta a_j$ | $\Delta \varphi_{j0}(°)$ |
| -5.76e-12 | 2.07e-12 | 1.22e-10 | -1.99e-13 | -2.00e-14 | 1.80e-12 |
| 3.41e-13 | 2.19e-13 | -7.68e-12 | 0 | 0 | -3.98e-13 |
| 9.54e-11 | **7.79e-10** | -2.23e-09 | -5.18e-09 | -4.23e-08 | 6.92e-08 |
| **1.99e-09** | -7.77e-10 | **-4.39e-08** | **6.41e-07** | **-2.51e-07** | **-6.72e-06** |
| 1.14e-13 | 4.26e-14 | -4.49e-12 | 3.50e-11 | -6.50e-13 | -4.22e-10 |
| 1.25e-12 | 4.68e-13 | -4.56e-11 | 2.20e-11 | -3.09e-12 | -2.09e-10 |
| -2.56e-13 | 9.70e-13 | -3.36e-12 | 9.95e-13 | -1.90e-13 | -6.20e-12 |
| -8.81e-13 | -8.34e-14 | 2.02e-11 | 0 | -2.00e-15 | 0 |
| 3.41e-13 | -9.43e-14 | -8.92e-12 | 0 | -1.30e-14 | 1.99e-12 |
| 3.41e-13 | -4.11e-15 | -7.67e-12 | 0 | -6.99e-15 | -4.23e-12 |

The color bold numbers are the maximum absolute errors of the component parameters. All maximum errors resulted from the two neighboring components with minimum frequency difference and maximum amplitude difference, and the other errors are greatly less.

Generally, the first-order and second-order derivatives of signals are unknown. In this case, they can be easily obtained by the differential circuits or numerically calculated by using the discrete values of the original signal.

**4.2 Simulation to FMCW radar measurement**

Frequency modulation continuous wave (FMCW) radar is widely used to measure the distance of targets. The transmitting antenna of the radar system transmits the FMCW radio signal, the receiving antenna receives the reflected signal from the target, and then mixing the transmission signal and reflection signal yields [31]

$$z(t) = Ae^{j(2\pi f_b t + \varphi_b)} + w(t) \tag{29}$$

where $A$ is the received signal power and $w(t)$ is the system noise. The frequency $f_b$ and phase $\varphi_b$ are given by the following equations

$$f_b = \frac{2BR}{cT_c} \tag{30}$$

$$\varphi_b = \frac{4\pi f_c R}{c} + \frac{4\pi BR^2}{c^2 T_c} \tag{31}$$

where $f_c$ is the chirp start frequency; $B$ is the bandwidth of the chirp; $T_c$ is the duration of the chirp; $R$ is the distance to the target; $c$ is the speed of light.

According to eqs. (23) and (24), the distance to the target can be calculated by measuring the frequency $f_b$ or the phase $\varphi_b$. In general, the distance measurement accuracy of the phase method is higher. However, because of the periodic ambiguity of the phase, when the measured distance is long, it is difficult to use the phase method [31]. Here, we simulate to measure the distance of the target by the frequency method and investigate the influence of noise on the measurement accuracy by using the proposed method of this paper.

Set the parameters of radar as follows: the chirp start frequency $f_c = 24\ GHz$, the bandwidth of the chirp $B = 100\ MHz$, the duration of the chirp $T_c = 512\ \mu s$, and the speed of light $c = 3 \times 10^8\ m/s$. Suppose the distance of the target $R = 12.3456789\ m$, and the noise $w(t)$ is Gaussian white noise with zero mean. With a specified signal-to-noise ratio (SNR), 1000 simulation measurements were made, and the simulated measurement results of distance were shown in figures 2 to 4.

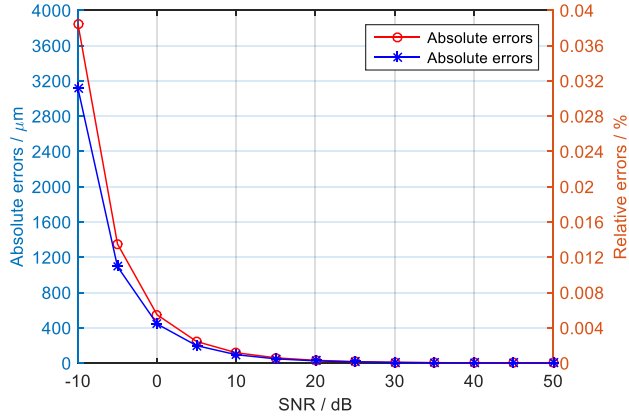

**Fig. 2.** The absolute errors and relative errors of the average distance of 1000 simulation measurements vs. SNR

Fig. 2 shows the absolute errors and relative errors of the average distance of 1000 simulation measurements vs. SNR. As seen from fig. 2, even if the SNR is very low, as low as -10 *dB*, the distance errors are less than 4 *mm,* and the relative error is approximately 0.032%. When SNR >= 10 *dB*, the absolute errors and relative errors are close to zero

The reason for the high distance accuracy is that the simulation measured frequency, amplitude and phase of the signal $z(t)$ are of high accuracy. Fig. 3 shows the comparison between the mean squared errors (MSEs) of the simulation measured frequency, amplitude and phase and their Cramer-Rao bounds (CRBs) [4].

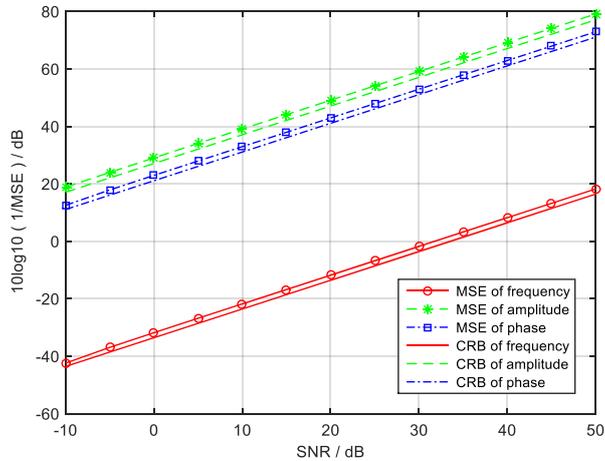

**Fig. 3.** The MSEs and CRBs of frequency, amplitude and phase of the signal vs. SNR. The MSEs of the frequency, amplitude and phase are less than those of their CRBs. This indicates that the method has achieved extraordinary accuracy.

In general theory, it is impossible that the MSEs of the frequency, amplitude and phase are less than their CRBs. The simulation results, however, show that the MSEs are indeed less than their CRBs. It is very difficult to explain these results at this time. On the hand, this indicates that the method has achieved extraordinary accuracy.

To investigate the robustness of the method, the MES of the simulation measured distance is calculated by the following equation

$$\sigma_s^2 = \frac{1}{N}\sum_{i=1}^{N}(R_i - R)^2 \qquad (32)$$

where $N = 1000$, is the simulation times; $R_i$ is the simulation measured distance at a certain SNR; $R$ is the real distance of target.

Fig. 4 gives the curve of $\sigma_s^2$ vs. SNR. It shows that $\sigma_s^2$ decreases rapidly with the increase of SNR. So the method has well robustness.

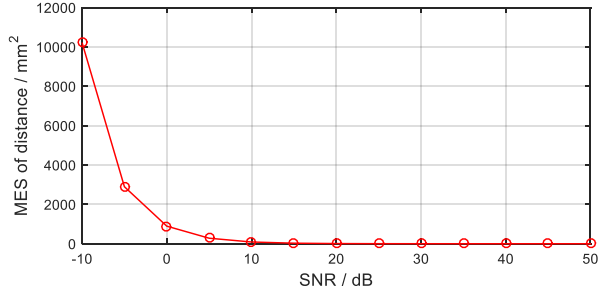

**Fig. 4.** The MSE of distance vs. SNR

## 5. Conclusions

The key point of this paper is the construction of generalized eigenvalue equations for real signals and complex signals. For the multifrequency real signal, the generalized eigenvalue equation is constructed by using the square Hankel matrixes of discrete values and their second-order derivatives of signal. For the multifrequency complex signal, the generalized eigenvalue equation is constructed by using the square Hankel matrixes of discrete values and their first-order derivatives of signal. The aforementioned Hankel matrix can be replaced by a Toeplitz matrix or other matrix as long as the rank of the matrix equals the rank of the corresponding square Hankel matrix. The first-order and second-order derivatives of signals can be obtained by differential circuits or numerically calculated by using the discrete values of the original signal.

The proposed two decomposition theorems are exactly in theory. For a noise-free real signal with *m* components, the proposed method can exactly compute the frequency, amplitude and phase of each component using only *2m*-1 discrete values and corresponding second-order derivatives. For a noise-free complex signal, the number of discrete values decreases to *2m*-1, and the second-order derivatives are replaced by the first-order derivatives. The proposed methods have very high resolution. The sampling rate is unnecessary to obey the sampling theorem.

With noisy signals, the proposed methods have extraordinary accuracy. It is possible that the MSEs of the frequency, amplitude and phase are less than their CRBs.

For the multifrequency discrete real signal and complex signal, it may be the first time to realize the exact decomposition of frequencies, amplitudes and phases of component signals in theory.

## Acknowledgment

The author would like to thank the financial support of the National Natural Science Foundation of China (Grant No. 12072106 & U1604254).

## References

1   Cooley, J. W. & Tukey J. W. An algorithm for the machine calculation of complex Fourier series. *Mathematics of Computation*, 19(90): 297-301, 1965.

2   Kumar, G. G., Sahoo, S. K. & Meher, P. K. 50 Years of FFT algorithms and applications. *Circuits Systems and Signal Processing*, 38(12): 5665-5698, 2019.

3   Weimann S., Perez-Leija A., Lebugle M., Keil R., Tichy M., Gräfe M., Heilmann R., Nolte S., Moya-Cessa H., Weihs G., Christodoulides D. N. & Szameit A. Implementation of quantum and classical discrete fractional fourier transforms. *Nature Communications*, 7: 11027, 2016.

4   Rife, D. C. & Vincent, G. A. Use of the discrete Fourier transform in the measurement of frequencies and levels of tones. *Bell Labs Technical Journal*, 49(2): 197-288, 1970.

5   Grandke, T. Interpolation algorithms for discrete Fourier transforms of weighted signals. *IEEE Transactions on Instrumentation and Measurement*, 32(2): 350-355, 1983.

6   Quinn, B. G. Estimating frequency by interpolation using Fourier coefficients. *IEEE Transactions on Signal Processing*, 42(5): 1264-1268, 1994.

7   Agrez, D. Weighted multipoint interpolated DFT to improve ampliture estimation of multifrequency signal. *IEEE Transactions on Instrumentation and Measurement*, 51(2), 287-292, 2002.

8   Aboutanios, E. & Mulgrew, B. Iterative frequency estimation by interpolation on Fourier coefficients. *IEEE Transactions on Signal Processing*, 53(4): 1237-1242, 2005.

9   Belega,D. & Petri, D. Accuracy analysis of the multicycle synchrophasor estimator provided by the interpolated DFT algorithm. *IEEE Transactions on Instrumentation and Measurement*, 62(5): 942-953, 2013.

10   Fan, L. & Qi, G. Frequency estimator of sinusoid based on interpolation of three DFT spectral lines. *Signal Processing*, 144: 52-60, 2018.

11   Tretter, S. A. Estimating the frequency of a noisy sinusoid by linear regression. *IEEE Transactions on Information Theory*, 31(6): 832-835, 1985.

12   Santamarida, I., Pantaleon, C. & Ibanez, J. A comparative study of high-accuracy frequency estimation methods. *Mechanical Systems and Signal Processing*, 14(5): 819-834, 2000.

13   Xiao, Y. C., Wei, P. & Xiao, X. C. Fast and accurate single frequency estimator. *Electronics Letters*, 40(14): 910-911, 2004.

14   Fu, H. & Kam, P. Y. Improved weighted phase averager for frequency estimation of single sinusoid in noise. *Electronics Letters*, 44(3): 247-248, 2008.

15   Liao, J. R. & Chen, C. M. Phase correction of discrete Fourier transform coefficients to reduce frequency estimation bias of single tone complex sinusoid. *Signal Processing*, 94: 108-117, 2014.

16   Kang, D., Ming X. & Xiaofei Z. Phase difference correction method for phase and frequency in spectral analysis. *Mechanical Systems and Signal Processing*, 14: 835-843, 2000.

17   Jacobsen, E. and Lyons, R. The sliding DFT. *IEEE Signal Processing Magazin*, 20(2): 74-78, 2003.

18   Wang, K., Zhang, L., Wen, H. and Xu, L. A sliding-window DFT based algorithm for parameter estimation of multi-frequency signal. *Digital Signal Processing*, 97: 102617, 2019.

19  Klema, V. & Laub, A. The singular value decomposition: Its computation and some applications. *IEEE Transactions on Automatic Control*, 25(2): 164-176, 1980.

20  Muruganatham, B., Sanjith, M. A. & Krishnakumar, B. Roller element bearing fault diagnosis using singular spectrum analysis. *Mechanical Systems and Signal Processing*, 35: 150-166, 2013.

21  Golafshan, R. & Sanliturk, K. Y. SVD and Hankel matrix based de-noising approach for ball bearing fault detection and its assessment using artificial faults. *Mechanical Systems and Signal Processing*, 70-71: 36-50, 2016.

22  Islam, M. T., Zabir, I. & Ahamed, S. T. A time-frequency domain approach of heart rate estimation from photoplethysmographic (PPG) signal. *Biomedical Signal Processing and Control*, 36: 146-154, 2017.

23  Grossmann, A. & Morlet, J. Decomposition of Hardy functions into square inte-grable wavelets of constant shape. *Siam Journal on Mathematical Analysis*, 15(4): 723-736, 1984.

24  Jiang, Q. & Suter, B. W. Instantaneous frequency estimation based on synchrosqueezing wavelet transform. *Signal Processing*, 138: 167-181, 2017.

25  Huang, N. E., Shen, Z. & Long, S. R. The empirical mode decomposition and the Hilbert spectrum for nonlinear and non-stationary time series analysis. *Proceedings of the Royal Society of London, Series A*, 454: 903-995, 1998.

26  Paulraj, A., Roy, R. & Kailath, T. A subspace rotation approach to signal parameter estimation. *Proceedings of the IEEE*, 74(7): 1044-1046, 1986.

27  Gu, Y. H. & Bollen, M. H. J. Estimating interharmonics by using sliding-window ESPRIT. *IEEE Transactions on Power Delivery*, 23(1): 13-23, 2008.

28  Odendaal, J. W., Barnard, E. & Pistorius, C. W. I. Two-dimensional super-resolution radar imaging using the MUSIC algorithm. *IEEE Transactions on Antennas and Propagation*, 42(10): 1386-1391, 1994.

29  Zheng, W., Li, X. L. & Zhu, J. Foetal heart rate estimation by empirical mode decomposition and MUSIC spectrum. *Biomedical Signal Processing and Control*, 42: 287-296, 2018.

30  Gallier, J. & Quaintance, J. Algebra, topology, differential calculus, and optimization theory for computer science and engineering. *Philadelphia*, PA 19104, USA, July 28, 2019, pp. 467–488.

31  Kim, B. S., Jin, Y., Lee, J. and Kim, S. High-Efficiency Super-Resolution FMCW Radar Algorithm Based on FFT Estimation. *Sensors*, 21(12): 4018, 2021.